\documentstyle[preprint,eqsecnum,aps]{revtex}
\begin{document}

\draft

\title{Surface Tension and Kinetic Coefficient for the
Normal/Superconducting Interface: Numerical Results vs.
Asymptotic Analysis}

\author{James C. Osborn\cite{jcoemail} and Alan T. Dorsey\cite{atdemail}}

\address{Department of Physics, University of Virginia\\
McCormick Road, Charlottesville, VA 22901}

\date{\today}

\maketitle

\begin{abstract}

The dynamics of the normal/superconducting interface in type-I
superconductors has recently been derived from the time-dependent
Ginzburg-Landau theory of superconductivity.   In a suitable limit
these equations are mapped onto a  ``free-boundary''
problem, in which the interfacial dynamics are determined by the
diffusion of magnetic flux in the normal phase. The magnetic field at
the interface satisfies a modified Gibbs-Thomson boundary condition
which involves both the surface tension of the interface and a kinetic
coefficient for motion of the interface.  In this paper we calculate
the surface tension and kinetic coefficient numerically by solving
the one dimensional equilibrium Ginzburg-Landau equations for a
wide range of $\kappa$ values.  We compare our numerical results to
asymptotic expansions valid for $\kappa\ll 1$,
$\kappa\approx 1/\sqrt{2}$, and $\kappa\gg 1$, in order to determine the
accuracy  of these expansions.

\end{abstract}



\section{Introduction}

When a type-I superconductor in a magnetic field is subjected to
either a sudden temperature or
magnetic field quench which takes it from the normal phase into the
Meissner phase, the approach to equilibrium will be determined
by the rate at which superconducting islands are nucleated in the
background normal phase, and the subsequent dynamics of the
superconducting/normal interfaces.  Refs. \cite{frahm91} and
\cite{liu91} suggested that the essential features
of the interface motion could be understood in terms of a free
boundary model for the magnetic field in the normal phase;
this model is almost identical to a free boundary model which
is used to study the growth of a solid into its supercooled
liquid phase.  The interface motion in the latter case is
known to be unstable, and leads to highly ramified solidification
patterns (dendrites, for instance; see Ref. \cite{kessler88} for an
overview).   By analogy, the growth of the
superconducting phase into the normal phase  should
be dynamically unstable.  Numerical
solutions of the time-dependent Ginzburg-Landau (TDGL) equations of
superconductivity confirmed these expectations \cite{frahm91,liu91}.
However, the precise connection between the TDGL equations and the
free boundary model was not made.

 More recently, the free boundary
model has been {\it derived} from the TDGL equations using the
method of matched asymptotic expansions \cite{chapman93,dorsey94}.
The free boundary model consists of a diffusion equation for the
magnetic field ${\bf h}$ in the normal phase,
\begin{equation}
\partial_{t} {\bf h} = D_{H} \nabla^{2} {\bf h},
\label{intro1}
\end{equation}
where $D_{H}=1/4\pi\sigma^{(n)}$ is the diffusion constant for the
magnetic flux, with $\sigma^{(n)}$ the normal state conductivity;
a continuity equation for the magnetic field at the normal/superconducting
interface,
\begin{equation}
(\nabla\times {\bf h})\times\hat{\bf n}|_{i} = -  D_{H} v_{n} {\bf h}_{i},
\label{intro2}
\end{equation}
with $v_{n}$ the interface velocity normal to the interface (with
$\hat{\bf n}\cdot {\bf h}_{i} = 0$); and a modified Gibbs-Thomson
boundary condition for the magnetic field at the interface,
\begin{equation}
h_{i} = H_{c} \left[ 1 - {4\pi\over H_{c}^{2}}\left( \sigma_{\rm ns}
 {\cal K} + \Gamma^{-1} v_{n} \right)\right],
\label{boundary6}
\end{equation}
where $H_{c}$ is the thermodynamic critical field for the superconductor,
$\sigma_{ns}$ is the surface tension of the
normal/superconducting interface, ${\cal K}$ is the curvature of the
interface, and $\Gamma^{-1}$ is a kinetic coefficient for motion of the
interface (here we ignore thermal fluctuations \cite{dorsey94}).
It is convenient to introduce the dimensionless
surface tension $\bar{\sigma}_{\rm ns}$ and kinetic coefficient
$\bar{\Gamma}^{-1}$, which are defined through
\begin{equation}
\sigma_{\rm ns} = {H_{c}^{2}\lambda \over 4\pi} \bar{\sigma}_{\rm ns},
\qquad \Gamma^{-1} = {H_{c}^{2}\lambda \over 4\pi }
        {2m\gamma \over \hbar \kappa^{2}}\bar{\Gamma}^{-1},
\label{dimensionless}
\end{equation}
where $\lambda$ is the magnetic penetration depth, $\kappa$ is the
Ginzburg-Landau parameter (the ratio of the penetration depth to the
coherence length $\xi$), $m$ is the mass of a Cooper pair, and
$\gamma$ is a dimensionless order parameter relaxation time.
The dimensionless surface tension and the kinetic coefficient can be
expressed in terms of
the solutions to the one dimensional {\it equilibrium} Ginzburg-Landau
equations, which are
\begin{equation}
{1\over \kappa^{2}} F'' - Q^{2} F + F - F^{3} = 0,
\label{GL1}
\end{equation}
\begin{equation}
Q'' - F^{2} Q= 0,
\label{GL2}
\end{equation}
where $F(x)$ is the dimensionless order parameter
amplitude and $Q(x)$  is the dimensionless magnetic vector potential
[the magnetic field is $H(x)=Q'(x)$]. For an interface between the
normal and superconducting phases, we have in the superconducting
phase $x\rightarrow -\infty$, $F(x)\rightarrow 1$ and $Q(x)\rightarrow  0$,
while in the normal phase $x\rightarrow\infty$, $F(x)\rightarrow 0$
and $Q(x)\sim x/\sqrt{2}$.  The surface tension and kinetic coefficient
are then \cite{chapman93,dorsey94}:
\begin{equation}
\bar{\sigma}_{\rm ns} = {1\over \kappa^{2}}\left[ I_{1}(\kappa)
              - I_{2}(\kappa)\right],
\label{tension}
\end{equation}
\begin{equation}
\bar{\Gamma}^{-1} = I_{1}(\kappa) - {2\pi\hbar\sigma^{(n)} \over m\gamma}
                                    I_{2}(\kappa),
\label{friction}
\end{equation}
where the integrals $I_{1}$ and $I_{2}$ are defined by
\begin{equation}
  I_{1}(\kappa) = 2 \int_{-\infty}^{\infty} dx\, (F')^{2}
  = 2\kappa^{2}\int_{-\infty}^{\infty}dx\,\left[ F^{2} - F^{4} -
F^{2}Q^{2}\right],
\label{I1}
\end{equation}
\begin{equation}
  I_{2}(\kappa) = 2\kappa^{2} \int_{-\infty}^{\infty} dx\,
    \left[ {1\over\sqrt{2}} Q' - (Q')^{2} \right]
       = 2\kappa^{2}\int_{-\infty}^{\infty} dx\, F^{2} Q^{2},
\label{I2}
\end{equation}
where second set of expressions are obtained by integrating by parts and
using Eqs.\ (\ref{GL1}) and (\ref{GL2}).
The Ginzburg-Landau equations have analytical solutions
only in certain limiting cases, to be discussed below.  Somewhat
surprisingly, there appear to be few numerical calculations of the
surface tension, even though the fundamental importance of this
quantity in distinguishing type-I and type-II superconductors was
recognized by Ginzburg and Landau in 1950 \cite{ginzburg50,numerical}.

In order to complete the derivation of the free boundary model
from the TDGL equations, in this paper we solve the equilibrium
Ginzburg-Landau equations numerically for a wide range
of $\kappa$ values, and use the solutions to calculate the
surface tension and kinetic coefficient.  Our paper is organized as follows.
In Sec. II we review
our numerical methods for solving the Ginzburg-Landau equations.
In Sec. III we  discuss the surface tension, and derive the asymptotic
form of the surface tension for small-$\kappa$,  which agrees well
with the numerical results for a large range of $\kappa$ values.
Our results also show that an asymptotic
expansion for the surface tension near  $\kappa=1/\sqrt{2}$, which
was derived in Appendix B of Ref. \cite{dorsey94}, has a larger
range of validity than expected. In Sec. IV we discuss our results for the
kinetic coefficient.  The kinetic coefficient can be either positive or
negative depending upon the ratio of relaxation times which appear
in the TDGL equations \cite{chapman93,dorsey94}; from our results we
are able to determine the range of parameter values which result in a
positive kinetic coefficient.  Sec. V is a discussion section in which
we briefly summarize our results.

\section{Numerical Methods}

Numerical computation is necessary to solve Eqs. (\ref{GL1}) and (\ref{GL2}) in
general. The method chosen was standard relaxation using Newton's method
\cite{recipes}. Since the relative size of the solution scales as $1/\kappa $
we rescaled the GL equations by substituting $x^{\prime }=\kappa x$
which results in the rescaled equations
\begin{equation}
 F'' - Q^{2} F + F - F^{3} = 0,
\label{GL1prime}
\end{equation}
\begin{equation}
\kappa^{2} Q'' - F^{2} Q= 0.
\label{GL2prime}
\end{equation}
The rescaled
differential equations can then be written as the following set of first
order finite difference equations:
\begin{eqnarray}
\Delta y_{1,k}/\Delta x_k^{\prime } &= \overline{y}_{3,k}, \nonumber\\
\Delta y_{2,k}/\Delta x_k^{\prime } &= \overline{y}_{4,k}, \nonumber \\
\Delta y_{3,k}/\Delta x_k^{\prime } &= \overline{y}_{1,k}\left[
  \left( \overline{y}_{2,k}\right) ^2+\left(
\overline{y}_{1,k}\right) ^2-1\right], \nonumber \\
\Delta y_{4,k}/\Delta x_k^{\prime } &=\left( \overline{y}_{1,k}\right) ^2
         \overline{y}_{2,k}/\kappa ^2,
\label{FDE}
\end{eqnarray}
with ${\bf y}_1=F$, ${\bf y}_2=Q$, ${\bf y}_3=F^{\prime }$,
${\bf y}_4=Q^{\prime}$, $\Delta y_{n,k}=y_{n,k}-y_{n,k-1}$,
$\Delta x_k^{\prime }=x_k^{\prime }-x_{k-1}^{\prime },$ and
$\overline{y}_{n,k}=\frac 12\left( y_{n,k}+y_{n,k-1}\right) $.
Each of the four equations is to be solved for $k=2\ldots M$, with $M$ the
number
of mesh points.
We also have four boundary conditions as follows:
\begin{eqnarray}
y_{1,1} &=1, \nonumber \\
y_{4,1} &=0, \nonumber \\
y_{1,M} &=0, \nonumber \\
y_{4,M} &=1/(\kappa \sqrt{2}),
\label{BNDRY}
\end{eqnarray}
giving a total of $4M$ equations of the $4M$ $y_{n,k}$'s. If we move the
right hand side of Eqs. (\ref{FDE}) to the left hand side
and multiply by $\Delta x_k^{\prime }$
we are then left with a set of homogeneous equations. Labeling these
equations $E_{n,k}$ for $n=1\ldots 4$, $k=2\ldots M$ the problem is
now to solve $E_{n,k}=0$ and Eq. (\ref{BNDRY}) simultaneously.
Given an initial guess ${\bf y}_k$
we can improve the solution using the expansion
\begin{equation}
{\bf E}_{k} ( {\bf y}_{k} + \Delta {\bf y}_{k}, {\bf y}_{k-1} + \Delta {\bf
y}_{k-1})
  \approx {\bf E}_{k} ( {\bf y}_{k}, {\bf y}_{k-1})
  + \sum_{n=1}^{4} {\partial {\bf E}_{k} \over \partial y_{n,k-1}}
  \Delta y_{n,k-1}
 + \sum_{n=1}^{4} {\partial {\bf E}_{k} \over \partial y_{n,k}} \Delta y_{n,k},
\label{NEWT}
\end{equation}
where we want the left hand side to equal zero.
This gives a system of linear equations
to solve for the $\Delta {\bf y}_k$'s. We then add the $\Delta {\bf y}_k$ to
the ${\bf y}_k$ to obtain a closer approximation. This process is repeated
until the maximum value of $\left|  E_{n,k}\right| $ is less than $%
10^{-6}$.

The mesh of points $x_k^{\prime }$ are chosen at the start of the
relaxation. We want the range to be large enough so that at the endpoints
the functions are already close to their values at infinity. By trying
different values for the endpoints we have found that at $x^{\prime }=\pm 100$
the functions are all sufficiently close to their boundary values at infinity
that
imposing the conditions (\ref{BNDRY}) here do not significantly affect the
solutions. We also want the mesh to be fine enough to accurately pick up the
detail of rapidly varying areas. Knowing that the solutions all show the
most change in a small area (taken to be around 0) and are relatively linear
outside, we have manufactured a grid with more closely spaced points
in the center than the edges using a total of 2001 points. The
spacing was chosen so as to make the
change in $y_1$ (i.e., $F$) from one mesh point to the other roughly constant.

Convergence for this algorithm depends greatly upon the initial guess. For $%
0.1<\kappa <10$ the solution is obtained within about 30 iterations from our
initial guess. Since we are interested in finding solutions for a large
range of $\kappa $ it is advantageous to use the previous solution to start
a new solution changing $\kappa $ slightly each time. In going from $\kappa =
$0.1 to 0.001 by 0.001 each subsequent solution was obtained in only three
iterations.

The results of our computations for large and small $\kappa$ are shown in
Figs. (\ref{fig1}) and (\ref{fig2}).  In Fig. (\ref{fig1}) $\kappa=10^{-3}$,
and we see that the magnetic field is essentially a step-function, as suggested
by the analysis in the next section.  In this case the field only
penetrates a short distance into the superconducting region, and therefore
the full positive energy of flux expulsion is obtained, resulting in a
positive surface tension (type-I superconductor).
In Fig. (\ref{fig2}) $\kappa=10$; we see
that the magnetic field penetrates far into the superconducting region,
so that the positive energy of flux expulsion is reduced.  However, the
negative condensation energy is very large here, resulting
in a net negative surface tension (type-II superconductor).
 From the numerical solutions we computed the surface tension and kinetic
coefficient using Eqs. (\ref{tension}), (\ref{friction}), (\ref{I1}),
and (\ref{I2}). Since our mesh spacing is
already adapted to rapidly varying parts of the solution, we used a basic
trapezoidal rule to calculate the integrals $I_{1}$ and $I_{2}$.  These
results are tabulated in the Table.  As expected, the surface tension
passes through zero at $\kappa = 1/\sqrt{2}$, separating type-I from
type-II superconductors.  In the following sections we will
compare these numerical results against asymptotic solutions of the
Ginzburg-Landau equations.

\section{Surface Tension}

The surface tension is the excess free energy per unit area due to
the presence of the interface.  As shown by Ginzburg and Landau
\cite{ginzburg50} (see also Ref. \cite{saintjames69}),
for $\kappa\ll 1/\sqrt{2}$,
$\bar{\sigma}_{\rm ns}= 2\sqrt{2}/3\kappa + O(\kappa^{-1/2})$.
Unfortunately, we have found that the lowest order expansion
provides a very poor approximation to the surface tension,
except for very small values of $\kappa$ (less than $10^{-3}$); this
fact was also noted by Ginzburg and Landau \cite{ginzburg50}.
Therefore, in this section we will first generalize their result somewhat by
calculating the next order term in the expansion.

In the small-$\kappa$ limit it is convenient to work with the
rescaled Ginzburg-Landau equations, Eqs.\ (\ref{GL1prime}) and
(\ref{GL2prime}).
The lowest order approximation is obtained by setting the first term
in the second Ginzburg-Landau equation, Eq.\ (\ref{GL2prime}),  equal to zero
so that $F^{2} Q= 0$.  In the superconducting phase  $Q=0$ with $F$ in the
superconducting phase determined by Eq.\ (\ref{GL1prime}) with
$Q=0$; the solution to this equation is $F(x')=-\tanh (x'/\sqrt{2})$,
for $x'>0$.
When this solution is substituted into the expression for the
surface tension it produces the lowest order expansion derived
by Ginzburg and Landau \cite{ginzburg50}.   To calculate the next order
term, we need to take this expression for the order parameter and
substitute it back into Eq.\ (\ref{GL2prime}), and then solve for
$Q$.  For $x'>0$ (the normal phase), we have $Q_{>}'' = 0$, which
integrates to $Q_{>}(x')  =x'/\kappa\sqrt{2} + C_{1}$, with $C_{1}$
a constant to be determined by matching onto the $x'<0$ solution.
For $x'<0$ (the superconducting phase), the vector potential satisfies
\begin{equation}
\kappa^{2} Q_{<}'' - \tanh^{2}(x'/\sqrt{2}) Q_{<}= 0.
\label{superconduct}
\end{equation}
Although this equation does not appear to have an explicit analytical
solution, for $\kappa\ll 1$ we can use the WKB method to obtain
an asymptotic solution.  Some care is necessary as this equation
has a second order turning point at $x'=0$.   The uniformly valid
asymptotic solution (i.e., a solution valid both near and
away from the turning point) is \cite{bender}
\begin{equation}
Q_{<}(x') = C_{2} { 2^{5/8} \Gamma (3/4) \over \pi} {1\over \kappa^{1/4}}
\left[ \ln\cosh(x'/\sqrt{2}) \over -\tanh (x'/\sqrt{2})\right]^{1/2}
 K_{-1/4}\left[ {\sqrt{2}\over \kappa} \ln\cosh (x'/\sqrt{2})\right],
\label{asympt}
\end{equation}
where $C_{2}$ is a second constant of integration and $K_{-1/4}(z)$ is the
modified Bessel function of order $-1/4$.  The integration
constants are determined by matching the solutions and their
derivatives at $x'=0$, with the result
\begin{equation}
C_{1}=C_{2} = - {\pi\over 2^{3/4} \Gamma(3/4)^{2}} {1\over \sqrt{\kappa}},
\label{constants}
\end{equation}
so that
\begin{equation}
Q_{<}(x') = - {1\over 2^{1/8} \Gamma(3/4) \kappa^{3/4} }
\left[ \ln\cosh(x'/\sqrt{2}) \over -\tanh (x'/\sqrt{2})\right]^{1/2}
 K_{-1/4}\left[ {\sqrt{2}\over \kappa} \ln\cosh (x'/\sqrt{2})\right].
\label{asympt2}
\end{equation}

To calculate the surface tension, we substitute our solution into
our expression for the surface tension, Eq.\ (\ref{tension}).
For $I_{1}$ we have
\begin{equation}
I_{1}(\kappa)  = 2\kappa \int_{-\infty}^{0} dx'\, \left[F^{2} -F^{4}
                            - F^{2}Q^{2} \right]
               = {2\sqrt{2} \kappa \over 3}
                 -{ 2^{3/4} \pi \over 8 \Gamma(3/4)^{2}} \kappa^{3/2} ,
\label{smallkappa1}
\end{equation}
and for $I_{2}$,
\begin{equation}
I_{2}(\kappa)  =  2\kappa \int_{-\infty}^{0}dx'\, F^{2} Q^{2}
           = { 2^{3/4} \pi \over 8 \Gamma(3/4)^{2}} \kappa^{3/2}.
\label{smallkappa2}
\end{equation}
Therefore, from Eq.\ (\ref{tension}) we find that the surface tension
in the small-$\kappa$ limit is
\begin{equation}
\bar{\sigma}_{\rm ns} = {2\sqrt{2} \over 3} {1\over \kappa}
                 -{ 2^{3/4} \pi \over 4 \Gamma(3/4)^{2}}
               {1\over \sqrt{\kappa}} + O(1).
\label{smallkappa3}
\end{equation}
The first term in the expansion was previously obtained by
Ginzburg and Landau \cite{ginzburg50}, and the second term is the
new result.  This calculation can also be formulated as a variational
calculation, with $F(x')=-\tanh (x'/\xi_{v}\sqrt{2})$ a trial
solution for the order parameter;  the WKB calculation may be
repeated, and the resulting solution used to calculate the surface tension
as a function of the variational parameter $\xi_{v}$.  Minimizing this
expression and taking the small-$\kappa$ limit, we obtain Eq.\
(\ref{smallkappa3})
\cite{lewis56}.

For $\kappa\gg 1/\sqrt{2}$ the second derivative term in the first
Ginzburg-Landau equation, Eq.\ (\ref{GL1}),  may be neglected, and the
resulting algebraic equation solved for $F$ as a function of $Q$.  This
expression is then substituted into the second Ginzburg-Landau equation,
Eq.\ (\ref{GL2}), and the resulting nonlinear differential equation
may also be solved \cite{saintjames69}.  The surface
tension in this limit is $\bar{\sigma}_{\rm ns} = -4(\sqrt{2} -1)/3$.

The surface tension is zero at $\kappa= 1/\sqrt{2}$ \cite{ginzburg50};
at this point the Ginzburg-Landau equations become
integrable \cite{dorsey94}.  The solutions may be used to carry out a
local analysis of the surface tension about $\kappa=1/\sqrt{2}$
\cite{dorsey94}, with the result
$\bar{\sigma}_{\rm ns} = 0.388 (1/2\kappa^{2} -1)$.

Summarizing, we  have
\begin{eqnarray}
\bar{\sigma}_{\rm ns} \approx \left\{
    \begin{array}{ll}
        0.943 \kappa^{-1}- 0.880\kappa^{-1/2},\qquad           \kappa\ll 1; \\
        0.388(1/2\kappa^{2} -1),\qquad      \kappa\approx 1/\sqrt{2} ; \\
        -0.552,\qquad            \kappa\gg 1.
    \end{array}
    \right.
\label{summary}
\end{eqnarray}

In Fig. (\ref{fig3}) the numerical results for the
surface tension are compared to the asymptotic expressions for $\kappa\ll 1$.
The asymptotic result is accurate for $\kappa <0.2$; the $\kappa^{-1/2}$
correction in Eq.\ (\ref{smallkappa3}) is important for values of
$\kappa$ which are greater that $10^{-3}$.   In Fig.\ (\ref{fig4}) we compare
the
numerical results against the asymptotic expansion derived in Ref.
\cite{dorsey94}
for $\kappa\approx 1/\sqrt{2}$.  The asymptotic expansion is reasonably
accurate for
$0.5<\kappa<1.0$.  If we identify the small parameter in this expansion to be
$\epsilon = 1/(2\kappa^{2}) -1$, then this would imply that the expansion is
accurate for values of $\epsilon$ as large as $\epsilon=1$, a somewhat
surprising result.  In this figure we also see that the surface tension changes
sign at $\kappa= 1/\sqrt{2}$, as expected.   Fig.\ (\ref{fig5}) shows the
surface tension in the range $1.0<\kappa<10.0$; for large $\kappa$ we see that
the surface tension is approaching the limiting value of $-0.55$.

\section{Kinetic Coefficient}

The kinetic coefficient $\bar{\Gamma}^{-1}$ is a function of $\kappa$,
which is a ratio of length scales, as well as $2\pi\hbar\sigma^{(n)}/ m\gamma$,
which is the ratio of the diffusion constant for the
order parameter, $D_{\psi} =  \hbar/2m\gamma$, to the diffusion
constant for the magnetic field,
$D_{H}= 1/4\pi\sigma^{(n)}$.  If this latter ratio is sufficiently large, the
kinetic coefficient may also change sign (resulting in some sort of dynamic
instability).  By setting $\bar{\Gamma}^{-1}=0$
in Eq.\ (\ref{friction}), we obtain an expression for the neutral
stability curve :
\begin{equation}
 \left[{m \gamma \over 2\pi\hbar\sigma^{(n)}}\right]_{\rm neutral} = {
I_{2}(\kappa)
                          \over I_{1}(\kappa)}.
\label{stability}
\end{equation}
The numerical result is plotted in Fig.\ (\ref{fig6}).

In the limit of small-$\kappa$ we may use the previously derived
expansions for $I_{1}$ and $I_{2}$ in Eqs.\ (\ref{smallkappa1})
and (\ref{smallkappa2}) to obtain
\begin{equation}
\bar{\Gamma}^{-1} = {2\sqrt{2}\over 3}\kappa -
\left( 1 + {2\pi\hbar\sigma^{(n)}\over m\gamma}\right)
 {2^{3/4}\pi \over 8 \Gamma(3/4)^{2}} \kappa^{3/2} + O(\kappa^{2}).
\label{smallkinetic}
\end{equation}
By setting $\bar{\Gamma}^{-1}=0$, we see that for small-$\kappa$ the stability
curve should behave as $\kappa^{1/2}$ for small-$\kappa$, which is confirmed
by the numerical results shown in Fig.\ (\ref{fig6}).

\section{Discussion}

We have investigated in some detail the behavior of the solutions to
the one-dimensional Ginzburg-Landau equations using both numerical methods
and asymptotic expansions.   The numerical results for the surface tension of
the
normal-superconducting interface agree well with the small-$\kappa$ asymptotic
expansion developed in this paper.  In addition, a recently developed
asymptotic expansion of the surface tension for
values of $\kappa$ near $1/\sqrt{2}$ \cite{dorsey94} agrees with the
numerical results over a surprisingly large range of $\kappa$ values.
We have also calculated the neutral stability curve for the kinetic
coefficient, which will be important in studies of the dynamics of
the normal-superconducting interface \cite{chapman93,dorsey94}.

\acknowledgments

This work was supported by NSF Grant  DMR 92-23586,
and by the Alfred P. Sloan Foundation.

\begin{figure}
\caption{Magnitude of the order parameter $F$  and the magnetic field $H$
for  $\kappa= 10^{-3}$, which corresponds to a type-I superconductor.
Lengths are in units of the penetration depth. }
\label{fig1}
\end{figure}

\begin{figure}
\caption{Magnitude of the order parameter $F$ and the magnetic field $H$
for  $\kappa= 10$, which corresponds to a type-II superconductor.
Lengths are in units of the penetration depth. }
\label{fig2}
\end{figure}

\begin{figure}
\caption{Dimensionless surface tension $\bar{\sigma}_{\rm ns}$ as a function
of the Ginzburg-Landau parameter $\kappa$ for $10^{-3}<\kappa<0.3$.
The solid line is the numerical result, and the dashed
line is the asymptotic expansion for $\kappa\ll 1$ given in
Eq.\ (\protect\ref{summary}).}
\label{fig3}
\end{figure}

\begin{figure}
\caption{Dimensionless surface tension $\bar{\sigma}_{\rm ns}$ as a function
of the Ginzburg-Landau parameter $\kappa$ for $0.3<\kappa<1.0$.
The solid line is the numerical result, and the dashed
line is the asymptotic expansion about $\kappa = 1/\protect\sqrt{2}$ given in
Eq.\ (\protect\ref{summary}).}
\label{fig4}
\end{figure}

\begin{figure}
\caption{Dimensionless surface tension $\bar{\sigma}_{\rm ns}$ as a function
of the Ginzburg-Landau parameter $\kappa$ for $1.0<\kappa<10.0$.
The solid line is the numerical result, and the dashed
line is the limiting value for $\kappa\gg 1$ given in
Eq.\ (\protect\ref{summary}).}
\label{fig5}
\end{figure}

\begin{figure}
\caption{Stability diagram for the kinetic coefficient, determined
by Eq.\ (\protect\ref{stability}) in the text. The $y$-axis is
 $m\gamma/2\pi\hbar\sigma^{(n)}$, the inverse of the
dimensionless conductivity.  For parameters in
the region above the line the kinetic coefficient is positive, while the region
below the line corresponds to a negative kinetic coefficient.}
\label{fig6}
\end{figure}

\begin{table}
\caption{Representative numerical results from the solution of
the Ginzburg-Landau equations.}
\label{table1}
\begin{tabular}{ddddd}
 $\kappa$ &   $I_{1}(\kappa)$ & $ I_{2}(\kappa)$ & $\bar{\sigma}_{\rm ns}$
    &    $\bar{\Gamma}^{-1}$ \tablenotemark[1] \\ \tableline

0.001       &  0.000926   &  0.0000161       &  910         &  0.000910  \\
0.01        &  0.00891    &  0.000516        &  84.0        &  0.00840   \\
0.05        &  0.0414     &  0.00586         &  14.2        &  0.0356   \\
0.1         &  0.0782     &  0.0169          &  6.13        &  0.0613   \\
0.2         &  0.144      &  0.0495          &  2.36        &  0.0942   \\
0.3         &  0.202      &  0.0943          &  1.19        &  0.107   \\
0.4         &  0.254      &  0.150           &  0.648       &  0.104   \\
0.5         &  0.301      &  0.217           &  0.338       &  0.0845   \\
0.6         &  0.345      &  0.294           &  0.142       &  0.0511   \\
$1/\sqrt{2}$ &  0.388      &  0.388          &  0.000219    &  0.000109   \\
1.0         &  0.490      &  0.706           &  -0.216      &  -0.216   \\
5.5         &  1.16       &  17.2            &  -0.529      &  -16.0   \\
10.0        &  1.43       &  55.8            &  -0.544      &  -54.4

\end{tabular}
\tablenotetext[1] {We have chosen $(2\pi\hbar\sigma^{(n)}/m\gamma)=1$ for the
purposes of illustration.}

\end{table}


\begin{references}
\bibitem{numerical} There is reference to some numerical work in
Ref. \cite{ginzburg50}, where it was determined
that for $\kappa=0.165$, $\bar{\sigma}_{\rm ns}\approx 3$.
\end{references}
\end{document}